\documentclass[prb,letterpaper,aps,floatfix,twocolumn]{revtex4-1}
\usepackage{graphicx}
\usepackage{amsmath}
\newcommand{\beq}{\begin{equation}}
\newcommand{\eeq}{\end{equation}}
\newcommand{\bk}{{{\bf{k}}}}

\newcommand{\bA}{{\bf{A}}}
\newcommand{\bB}{{\bf{B}}}

\newcommand{\bb}{{\bf{b}}}

\newcommand{\beqa}{\begin{eqnarray}}
\newcommand{\eeqa}{\end{eqnarray}}

\newcommand{\dg}{{\dag}}
\newcommand{\pdg}{{\vphantom \dag}}
\newcommand{\btau}{{\boldsymbol \tau}}
\newcommand{\bsigma}{{\boldsymbol \sigma}}

\newcommand{\bnabla}{{\boldsymbol \nabla}}
\newcommand{\bpi}{{\boldsymbol \pi}}
\begin{document}
\title{Weyl semimetal with broken time reversal and inversion symmetries}
\author{A.A. Zyuzin}
\author{Si Wu}
\author{A.A. Burkov}
\affiliation{Department of Physics and Astronomy, University of Waterloo, Waterloo, Ontario 
N2L 3G1, Canada}
\date{\today}
\begin{abstract}
Weyl semimetal is a new topological state of matter, characterized by the presence of nondegenerate band-touching nodes, 
separated in momentum space, in its bandstructure. 
Here we discuss a particular realization of a Weyl semimetal: a superlattice heterostructure, made of alternating layers 
of topological insulator (TI) and normal insulator (NI) material, introduced by one of us before.  
The Weyl node splitting is achieved most easily in this system by breaking time-reversal (TR) symmetry, for example by magnetic doping.  
If, however, spatial inversion (I) symmetry remains, 
the Weyl nodes will occur at the same energy, making it possible to align the Fermi energy simultaneously with both nodes. 
The goal of this work is to explore the consequences of breaking the I symmetry in this system. We demonstrate that, while 
this generally moves the Weyl nodes to different energies, thus eliminating nodal semimetal and producing a state with electron and
hole Fermi surfaces, the topological properties of the Weyl semimetal state, i.e. the chiral edge states and the corresponding Hall conductivity,
 survive for moderate I symmetry breaking. Moreover, we demonstrate that a new topological phenomenon arises in this case, if an external magnetic 
 field along the growth direction of the heterostructure is applied. Namely, this leads to an equilibrium dissipationless current, flowing along 
 the direction of the field, whose magnitude is proportional to the energy difference between the Weyl nodes and to the magnetic field, with a universal 
 coefficient, given by a combination of fundamental constants. 
\end{abstract}
\maketitle
\section{Introduction}
An interesting recent development in the study of topological phases in condensed matter is the 
realization that such phases can exist in {\em gapless semimetals}.~\cite{Vishwanath11,Balents11,Ran11,Burkov11-1,Burkov11-2,Kim11,Fang11,Halasz11,Hosur11,Aji11} 
This goes against the common wisdom that a bulk spectral gap is a necessary ingredient 
in any topological phase, eliminating the sensitivity to small perturbations that any gapless phase would 
naively exhibit.   
Such {\em topological semimetals} are characterized by the presence of two or more nondegenerate accidental band-touching 
points in their bandstructure. 
As first understood by Volovik,~\cite{Volovik03,Volovik07,Volovik11} these points are characterized by a nontrivial topology, 
and can be thought of as hedgehog-like topological defects and point sources of Berry flux in momentum space, characterized by 
an integer ($\pm 1$) topological charge. 
The $k \cdot p$ Hamiltonian near such band-touching points takes the form of $2 \times 2$ Hamiltonian of a chiral Weyl 
fermion, hence the name {\em Weyl semimetal}.

In the presence of both time-reversal (TR) and inversion (I) symmetries, all bands are doubly degenerate at all momenta in the Brillouin zone (BZ), as follows from the 
Kramers theorem. 
This means that Weyl semimetals require breaking of either TR or I symmetries,~\cite{Murakami} which makes a contact between only two bands possible, separating the individual Weyl 
nodes in momentum space. 
When Weyl nodes are separated from each other in momentum space, they can not be hybridized, which makes them indestructible, as they can only 
disappear by mutual annihilation of pairs with opposite topological charges. 
This is the mechanism of topological stability of a Weyl semimetal state, which is distinct from the spectral-gap protection in insulating topological phases.~\cite{footnote1} 

Perhaps the most interesting variety of Weyl semimetals is obtained when TR is broken. As first shown in Ref.~\onlinecite{Burkov11-1}, 
the simplest kind of a Weyl semimetal can be realized in this case, with only two, the minimal number required by the fermion-doubling 
theorem,~\cite{Nielsen81} Weyl band-touching nodes. This is achieved in a magnetically-doped multilayer heterostructure, consisting of 
thin layers of topological insulator (TI) material, with normal insulator (NI) spacer layers in between. 
This particular realization of a Weyl semimetal is characterized by a semi-quantized anomalous Hall (AH) conductivity, proportional 
to the separation between the Weyl nodes in momentum space, and chiral surface states, and can be thought of as the closest three-dimensional (3D)
analog of a 2D integer quantum Hall state. We will limit our discussion to this system henceforth. 

While the broken TR is necessary to separate the Weyl nodes in momentum space and create a topologically-stable phase, the remaining I symmetry also
plays an important role. This symmetry guarantees that the two Weyl nodes, separated along the growth direction of the TI-NI multilayer in momentum space, occur at the same energy. 
This makes it possible for the Fermi energy to coincide with both Weyl nodes simultaneously and thus realize a true nodal semimetal phase. 
The goal of this paper is to explore the consequences of breaking the inversion symmetry in the TI-NI multilayer Weyl semimetal, in addition to breaking TR.    
As we demonstrate, a moderate breaking of inversion symmetry does not destroy the topological nature of the state, even though it generally moves Weyl nodes 
to different energies, thus destroying the {\em nodal semimetal} and producing a state with electron and hole Fermi surfaces. However, topological properties survive, as 
long as the I breaking is not too strong. Moreover, we demonstrate that a new topological phenomenon arises in the case, when the Weyl nodes are separated 
in both energy and momentum. Namely, an external magnetic field, applied along growth direction of the heterostructure, 
produces a nondissipative ground state 
current. This current is associated exclusively with the zero-mode Landau levels (LLs) of the system and has a topological character, in the sense made precise below. 

The rest of paper is organized as follows. In section~\ref{sec:2} we present a general Chern-Simons-like description of a Weyl semimetal, which demonstrates, that in addition to 
the anomalous Hall effect, there exists another topological effect, namely the magnetic-field driven equilibrium current, which arises when the Weyl nodes are separated not 
only in momentum space, but also in energy. In section~\ref{sec:3} we present a model of a TI-NI multilayer heterostructure with broken TR {\em and} inversion symmetries and 
discuss the general phase diagram of this system. In section~\ref{sec:4} we present an explicit linear-response calculation of the magnetic-field driven equilibrium current in the 
TI-NI multilayer. We conclude in section~\ref{sec:5} with a discussion of the experimental observability of the proposed effect. 

\section{Chern-Simons-like description of the Weyl semimetal}
\label{sec:2}
We will start with a general discussion of the topological response in Weyl semimetals, based on $3+1$-dimensional Chern-Simons-like theory, 
which nicely unifies both the semi-quantized AH response, discussed before,~\cite{Burkov11-1,Burkov11-2} and the new topological magnetic field-induced 
current effect, proposed in this paper. 
The fact that chiral Weyl fermions, separated in momentum and energy, induce a Chern-Simons-like term in the action of the electromagnetic 
field, is well-known in the particle-physics literature,~\cite{Jackiw90,Jackiw99,Perez99,Volovik05} where it arises in the context of CPT and Lorentz-violating 
extensions of the Standard Model. The action, written in the real time representation, has the following form:
\beq
\label{eq:1}
S_{CS} = \frac{1}{2} \int d^4 x\,\, b^{\mu} \epsilon_{\mu \alpha \beta \gamma} A^{\alpha} \partial_{\beta} A^{\gamma},  
\eeq
where $A^{\mu}$ is the electromagnetic gauge potential and $\mu = 0,1,2,3$. 
Here we use the standard notation for the four-vectors $x^{\mu} = (x^0, {\bf x})$, $x_{\mu} = (x^0, - {\bf x})$, where $x^0$ is the 
temporal component. 
The four-vector $b^{\mu}$ selects a direction in the 4-dimensional space-time, thus violating Lorentz invariance. 
Its temporal component, $b^0$, is proportional to the energy difference between the Weyl nodes (we restrict ourselves to the case of only two Weyl nodes), 
while its spatial component, $\bb$, is proportional to the momentum-space separation between them.~\cite{Jackiw99,Perez99,Volovik05}
Let us first assume that $b^{\mu}$ has no temporal component. By a spatial rotation we can then bring it to the form $b^{\mu} = (0,0,0,b)$, 
which corresponds to Weyl nodes, separated along the $z$-direction in momentum space. 
As shown in Ref.~\onlinecite{Burkov11-1}, such a state is characterized by a nonzero Hall conductivity, given by:
\beq
\label{eq:2}
\sigma_{xy} = \frac{e^2 k_0}{\pi h}, 
\eeq
where $2 k_0$ is the separation between the Weyl nodes in momentum space. 
This result can be obtained from the Chern-Simons action Eq.~\eqref{eq:1}. Indeed the current density can be calculated as:
\beq
\label{eq:3}
j_{\alpha} =  \frac{\delta S_{CS}}{\delta A^{\alpha}} = b^{\mu} \epsilon_{\mu \alpha \beta \gamma} \partial_{\beta} A^{\gamma},
\eeq
which gives:
\beq
\label{eq:4}
j_x = - b \left( \frac{\partial \varphi}{\partial y} + \frac{1}{c} \frac{\partial A^y}{\partial t} \right)  = b E^y, 
\eeq
where $\varphi = A^0$. 
Thus in this case $b = \sigma_{xy}$. 
 
Now let us take $b^{\mu}$ to be purely time-like, i.e. $b^{\mu} = (b, 0, 0,0)$. 
Taking a functional derivative of the Chern-Simons-like action with respect to the vector potential, as above, we obtain:
\beq
\label{eq:5}
j_{\alpha} = b \epsilon_{0 \alpha \beta \gamma} \partial_{\beta} A^{\gamma}.
\eeq
Taking $\alpha = 3$, we then obtain;
\beq
\label{eq:6}
j_z = - j_3 = - b \epsilon_{0 3 \beta \gamma} \partial_{\beta} A^{\gamma}  = - b B_z, 
\eeq
where $\bB = \bnabla \times \bA$ is the magnetic field. 
The physical meaning of this result is that a magnetic field, applied to a Weyl semimetal with an energy separation between the nodes, 
will induce a current in the direction of the applied field. This current exists in equilibrium and is thus nondissipative.   
This result is known in the particle physics literature,~\cite{Vilenkin80,Kharzeev08} in the context of parity-violating models of massless chiral particles, as
{\em chiral magnetic effect}.  
As we will demonstrate below, it is realized in a simple model of a Weyl semimetal with broken TR and I symmetries.  

A couple of comments are in order here. First, it may appear from the above discussion, that energy separation between the Weyl nodes is enough in order 
to obtain the dissipationless current and no momentum space separation is needed. While this is true in an abstract phenomenological model, in which Weyl fermions of 
opposite chirality are simply introduced by hand, it is certainly not true in our microscopic model, where Weyl fermions appear only when separated in momentum space, except 
at a single point in parameter space, corresponding to the TI-NI transition in the absence of TR and I breaking. Correspondingly, as will be shown below, the time-like parameter $b$ in the 
Chern-Simons theory  
is only nonzero when the Weyl nodes are separated in momentum space, except if the multilayer system is fine-tuned to the TI-NI transition point in the absence of TR-breaking 
[see Eq.~\eqref{eq:47} below]. Second, Eq.~\eqref{eq:5} is explicitly isotropic in space, i.e. the magnetic field and the corresponding dissipationless current can have any direction. 
Our microscopic model, on the other hand, is anisotropic (there is a preferred direction---the growth direction of the multilayer) and only becomes effectively isotropic at low energies, long distances 
and weak applied fields. Thus, in all our calculations below we assume that the magnetic field is applied along the growth direction of the multilayer, since in this case all the calculations can be done exactly, without 
making any additional assumptions of weak fields and low energies. The dissipationless current, however, will exist if the magnetic field is applied along any other direction, in accordance with Eq.~\eqref{eq:5}.   

\section{TI-NI multilayer with broken time-reversal and inversion symmetries}
\label{sec:3}
We consider a realization of a Weyl semimetal, based on a magnetically-doped TI-NI multilayer heterostructure, introduced by one of us in Ref.~\onlinecite{Burkov11-1}:
\beqa
\label{eq:7}
H&=&\sum_{\bk_{\perp}, ij} \left[ v_F \tau^z (\hat z \times \bsigma) \cdot \bk_{\perp} \delta_{i,j} + m \sigma^z \delta_{i,j}
+ \Delta_S \tau^x \delta_{i,j} \right. \nonumber \\ 
&+& \left.\frac{1}{2} \Delta_D \tau^+ \delta_{j, i+1} + \frac{1}{2} \Delta_D \tau^- \delta_{j, i-1} \right] c^\dg_{\bk_{\perp} i} c^\pdg_{\bk_{\perp} j}. 
\eeqa
The first term in Eq.(\ref{eq:7}) describes the two (top and bottom) surface states of an individual TI layer. 
$v_F$ is the Fermi velocity, characterizing the surface Dirac fermion, which we take to be the same on the top and 
bottom surfaces of each layer. $\bk_{\perp}$ is the momentum in the 2D surface BZ (we use $\hbar = 1$ units), 
$\bsigma$ is the triplet of Pauli matrices, acting on the real spin degree of freedom, and $\btau$ are Pauli matrices, acting 
on the {\em which surface} pseudospin degree of freedom. The indices $i,j$ label distinct TI layers. 
The second term describes exchange spin splitting of the surface states, which can be induced, for example, by doping 
each TI layer with magnetic impurities.~\cite{Chen10,Chang11}
The remaining terms in Eq.(\ref{eq:7}) describe tunneling between top and bottom surfaces within the same TI layer (the term, proportional 
to $\Delta_S$), and between top and bottom surfaces of neighboring TI layers (terms, proportional to $\Delta_D$).  
Longer-range tunneling is assumed to be negligible. 
Partially diagonalizing Eq.~\eqref{eq:7} by Fourier transform, we obtain a momentum space Hamiltonian:
\beq
\label{eq:7a}
{\cal H}(\bk) = v_F \tau^z (\hat z \times \bsigma) \cdot \bk + m \sigma^z + \hat \Delta(k_z), 
\eeq
where $\hat \Delta(k_z) = \Delta_S \tau^x +  \frac{1}{2} (\Delta_D \tau^+ e^{i k_z d} + H.c. )$. 

The TI-NI multilayer, described by Eq.~\eqref{eq:7a}, possesses inversion symmetry with respect to an inversion center placed midway between 
the top and bottom surfaces in any TI or NI layer. Explicitly, the inversion operation changes the sign of the momentum and interchanges the top and bottom surfaces in each TI layer (spin $\bsigma$ is 
a pseudovector and is thus invariant under inversion):
\beq
\label{eq:7b}
{\cal I}: {\cal H}(\bk) \rightarrow \tau^x {\cal H}(- \bk) \tau^x.
\eeq
It is easy to see that Eq.~\eqref{eq:7a} is indeed invariant under ${\cal I}$. 

We now relax the assumption that the multilayer is inversion-symmetric, which is perhaps more realistic, 
as some degree of structural inversion asymmetry will be inevitably introduced during growth. Note that inversion-asymmetric TI-NI multilayer 
model (but with preserved TR symmetry) was considered before in Ref.~\onlinecite{Halasz11}. 

We thus need to add terms to the multilayer Hamiltonian Eq.~\eqref{eq:7a}, which violate inversion symmetry. There are of course infinitely many such terms
one can write down and we will restrict ourselves only to terms, independent of momentum (i.e. leading-order terms, arising in the expansion of a general I-breaking 
term with respect to the momentum).
A simple analysis shows that there are only two such terms: $V \tau^z$ and $\lambda \tau^y \sigma^z$, which break inversion, but respect all other symmetries of the multilayer, 
in particular rotation around the $z$-axis.  The first term corresponds to an 
electrostatic potential difference between the top and bottom surface of a TI layer in each unit cell (note that the total electric field in each unit cell vanishes). 
The second term is momentum-independent spin-orbit interaction term, which is allowed by the broken inversion symmetry. This term is closely analogous to the Dresselhaus 
spin-orbit interaction terms in semiconductors lacking inversion symmetry.~\cite{Dresselhaus} It is simpler than, say, the well-known Dresselhaus terms in semiconductors
with zinc-blende structure due to the fact that our system has a uniaxial anisotropy.  

To understand the effect of the I-breaking terms on the physical properties of the multilayer, it is useful to start from the case when neither TR nor I are violated. 
In this case, diagonalizing Eq.~\eqref{eq:7a} gives four bands, which are pairwise-degenerate at every momentum, as required by the Kramers theorem:
\beq
\label{eq:8}
\epsilon_{\pm}(\bk) = \pm \sqrt{v_F^2(k_x^2 + k_y^2) + \Delta^2(k_z)}, 
\eeq
where $\Delta(k_z) = \sqrt{\Delta_S^2 + \Delta_D^2 + 2 \Delta_S \Delta_D \cos(k_z d)}$ and $d$ is the superlattice period. 
Let us now break I without breaking TR, i.e. add the $V \tau^z$ and $\lambda \tau^y \sigma^z$ terms to the Hamiltonian, but keep $m = 0$. 
In this case the band dispersion can still be found analytically and one obtains:
\beqa
\label{eq:9}
&&\epsilon^2_{\pm}(\bk) = v_F^2 (k_x^2 + k_y^2) + V^2 + \lambda^2 + \Delta^2(k_z) \nonumber \\
&\pm&2 \sqrt{v_F^2 (V^2 + \lambda^2) (k_x^2 + k_y^2) + \frac{1}{2} \lambda^2 \Delta_D^2 [1 - \cos(2 k_z d)]}. \nonumber \\
\eeqa
Thus the double degeneracy at every momentum is lifted, as it should be, except at the I and TR-invariant momenta $k_x = k_y = 0$ and $k_z = 0, \pi/d$. 
The role of the $V \tau^z$ term is to split the degeneracy for $k_x, k_y \ne 0$, while the role of the $\lambda \tau^y \sigma^z$ term is 
to split the degeneracy at $k_x = k_y = 0$ and $k_z \ne 0, \pi/d$. In other words, in the absence of the $\lambda \tau^y \sigma^z$ term, the double degeneracy 
remains everywhere along the line $k_x = k_y = 0$ in momentum space. 

Let us now also break the TR symmetry and add the spin-splitting term $m \sigma^z$. In general, it is now impossible to find the energy eigenvalues of the multilayer 
Hamiltonian analytically, except in a few special cases, which we will now discuss, before considering the most general situation. 
The first special case we will consider is $\lambda = 0$. The band dispersion in this case is given by:
\beqa
\label{eq:10}
\epsilon^2_{\pm}(\bk)&=&v_F^2 (k_x^2 + k_y^2) + V^2 + m^2 + \Delta^2(k_z) \nonumber \\
&\pm&2 \sqrt{v_F^2 V^2  (k_x^2 + k_y^2) + m^2 V^2 + m^2 \Delta^2(k_z)}. \nonumber \\
\eeqa
The $\epsilon_-(\bk)$ pair of bands exhibits band-touching nodes, which are solutions of the equation:
\beq
\label{eq:11}
\left[ v_F^2(k_x^2 + k_y^2) + m^2 + \Delta^2(k_z) - V^2 \right]^2 = 4 (m^2 - V^2) \Delta^2(k_z). 
\eeq
Clearly, two cases must be considered: $m > V$ and $m < V$. In the latter case, a solution exists only when $\Delta(k_z) = 0$, which happens 
when $\Delta_S/\Delta_D = \pm 1$ at $k_z = \pi/d, 0$ correspondingly. Eq.~\eqref{eq:11} simplifies to:
\beq
\label{eq:12}
v_F^2 (k_x^2 + k_y^2) = V^2 - m^2. 
\eeq
This describes a circular nodal line in the $xy$-plane, which, however, exists only at the TI-NI critical point $\Delta_S/\Delta_D = \pm 1$ (a stable Weyl semimetal phase 
can be obtained in this case, if one includes momentum dependence of the $\Delta_{S,D}$ tunneling amplitudes, breaking the $z$-axis rotational symmetry;
we will not pursue this direction here, as it has been investigated in detail in Ref.~\onlinecite{Halasz11}).   

In the $m > V$ case, Eq.~\eqref{eq:11} simplifies to:
\beq
\label{eq:13}
v_F^2 (k_x^2 + k_y^2) + \left[ \sqrt{m^2 - V^2} - \Delta(k_z) \right]^2 = 0. 
\eeq
The solution is:
\beq
\label{eq:14}
k_x = k_y = 0, \,\, \Delta(k_z) = \sqrt{m^2 - V^2}. 
\eeq
This corresponds to the point-node Weyl semimetal, considered in Ref.~\onlinecite{Burkov11-1}. 
The effect of the $V \tau^z$ term in this case is only to reduce the distance between the Weyl nodes in momentum space. 
Henceforth, we will assume that $m > V$ and, keeping the above analysis in mind, neglect the $V \tau^z$ term entirely. 

Now let us analyze the effect of the I-breaking spin-orbit interaction term $\lambda \tau^y \sigma^z$ in the presence of 
the broken TR symmetry due to the $m \sigma^z$ term. In this case, the band dispersion can not be found analytically in general, 
except at the $k_x = k_y = 0$ point. At this point we obtain:
\beq
\label{eq:15}
\epsilon_{s \pm} (k_z) = m s \pm \sqrt{\lambda^2 + \Delta^2(k_z) - 2 s \lambda \Delta_D \sin(k_z d)},
\eeq
where $s = \pm$. 
The Weyl nodes are given by the solution of the equation:
\beq
\label{eq:16}
\epsilon_{+ -}(k_z) = \epsilon_{- +}(k_z), 
\eeq
or, explicitly:
\beqa
\label{eq:17}
&&2m = \sqrt{\lambda^2 + \Delta^2(k_z) - 2 \lambda \Delta_D \sin(k_z d)} \nonumber \\
&+&\sqrt{\lambda^2 + \Delta^2(k_z) + 2 \lambda \Delta_D \sin(k_z d)}. 
\eeqa
It is obvious already from the form of Eq.~\eqref{eq:17} that the nodes no longer occur at zero energy. 
Eq.~\eqref{eq:17} can not be solved analytically in its general form, but we can solve it perturbatively in $\lambda$. 
To first order in $\lambda$, the position of the nodes along the $z$-axis in momentum space in unchanged and is 
given by the equation:
\beq
\label{eq:18}
m = \Delta(k_z). 
\eeq
Assuming for concreteness that the tunneling amplitudes $\Delta_{S,D}$ are both positive, the nodes are then located at $k^{\pm}_z = \pi/d \pm k_0$, where:~\cite{Burkov11-1}
\beq
\label{eq:19}
k_0 = \frac{1}{d}\textrm{arccos}\left(\frac{\Delta_S^2 + \Delta_D^2 - m^2}{2 \Delta_S \Delta_D} \right), 
\eeq
The shift of the Weyl nodes from the zero energy can then be obtained from Eq.~\eqref{eq:15} by expanding it to first order in $\lambda$. We obtain:
\beqa
\label{eq:20}
\Delta \epsilon_{\pm}&=&\left.  \frac{\lambda \Delta_D \sin(k_z d)}{\Delta(k_z)}\right|_{k_z^{\pm} = \pi/d \pm k_0} \nonumber \\
&=&\mp \frac{\lambda}{2 \Delta_S m} \sqrt{(m_{c2}^2 - m^2) (m^2 - m_{c1}^2)}, 
\eeqa
where $m_{c1} = |\Delta_S - \Delta_D|$ and $m_{c2} = \Delta_S + \Delta_D$ are, correspondingly, the lower and upper critical values of the 
spin splitting $m$, between which the Weyl semimetal phase exists.~\cite{Burkov11-1}

\begin{figure}[t]
\includegraphics[width=8cm]{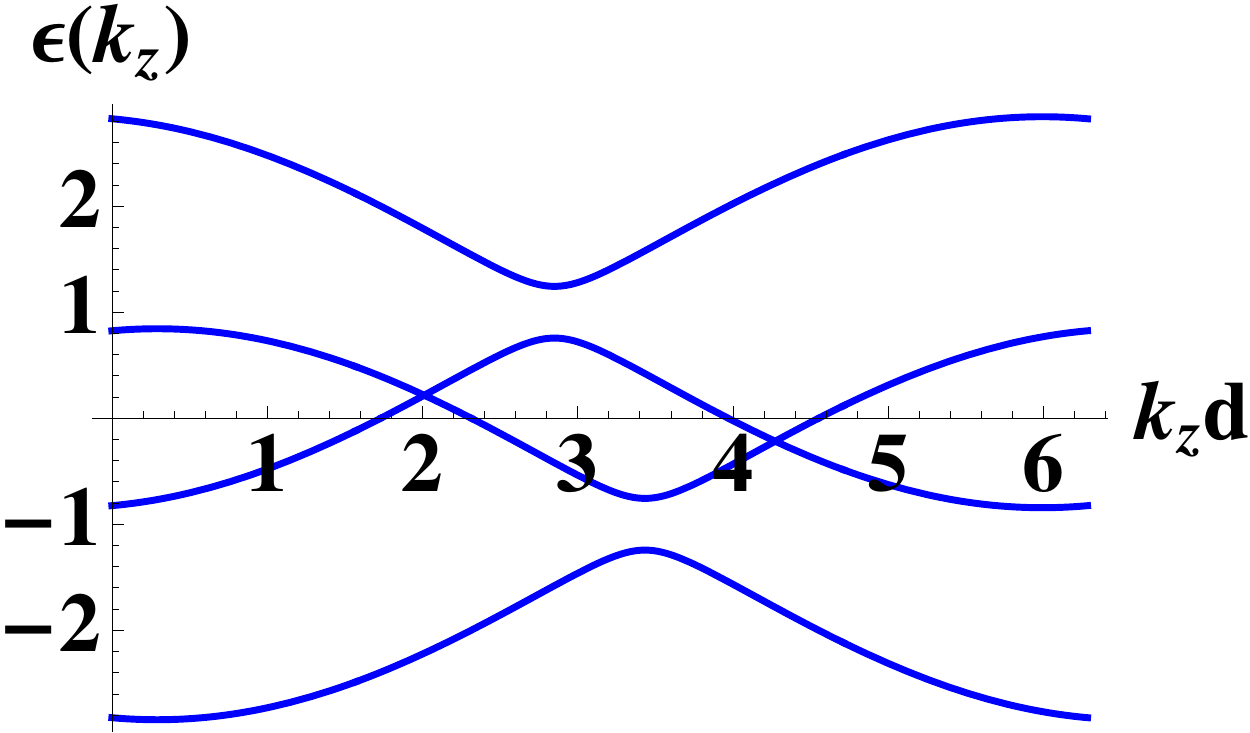}
\caption{(Color online) A plot of the band dispersion along the $k_x = k_y =0$ line for a moderate magnitude of the inversion symmetry breaking, corresponding 
to $\lambda/\Delta_S = 0.3$. The Weyl band-touching nodes are visibly shifted in opposite directions in energy, but are otherwise intact. At charge neutrality the Fermi energy 
is pinned at $\epsilon_F = 0$, and there are equal-volume electron and hole Fermi surfaces.}
\label{fig:1}
\end{figure}

\begin{figure}[t]
\includegraphics[width=8cm]{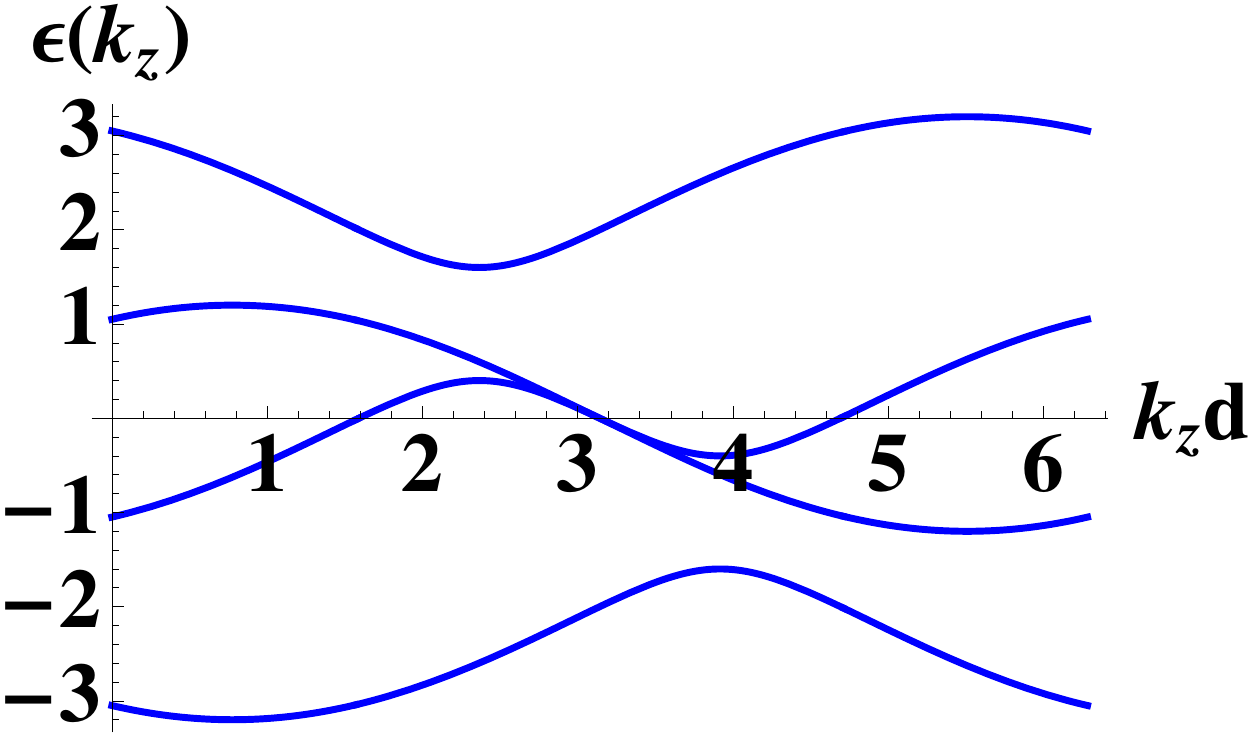}
\caption{(Color online) Band dispersion along the $k_x = k_y =0$ line for $\lambda = \lambda_{c1} = \sqrt{m^2 - (\Delta_S - \Delta_D)^2}$. The electron and hole Fermi surfaces touch and the Weyl nodes
disappear.}
\label{fig:2}
\end{figure}

\begin{figure}[t]
\includegraphics[width=8cm]{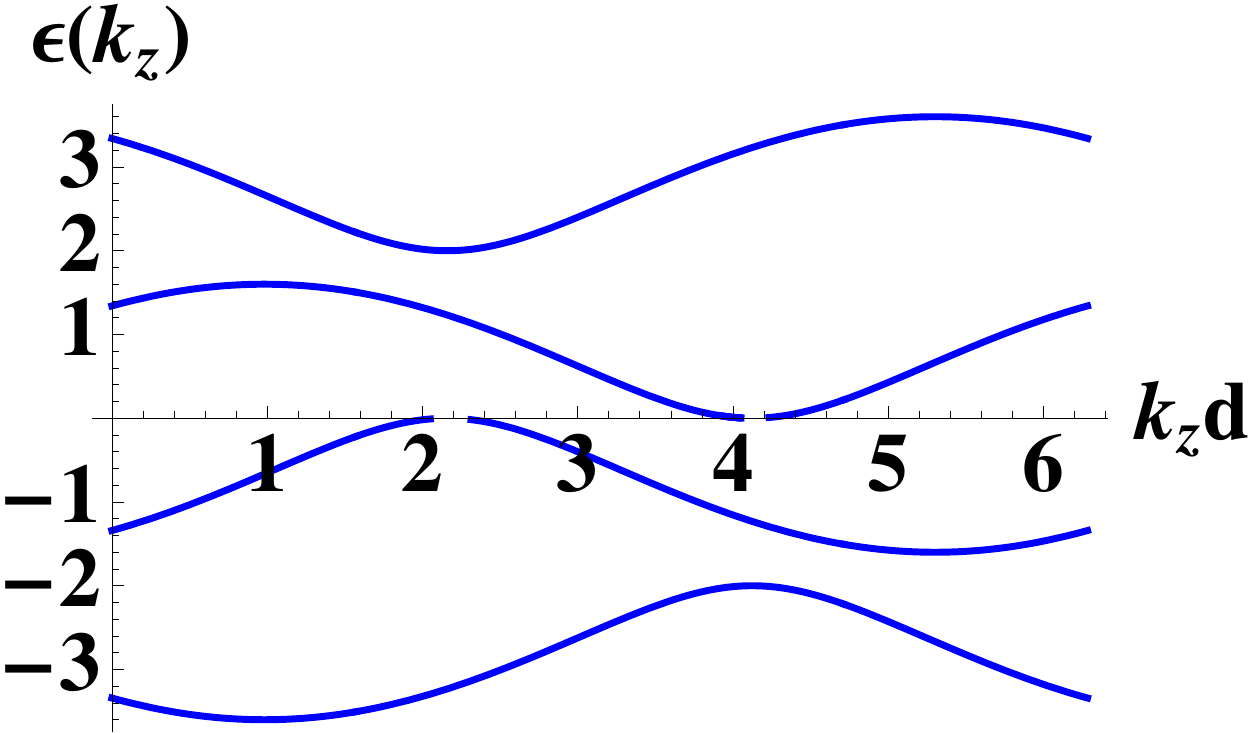}
\caption{(Color online) Band dispersion along the $k_x = k_y =0$ line for $\lambda = \lambda_{c2} = \sqrt{(m + \Delta_D)^2 - \Delta_S^2}$. The Fermi surfaces disappear entirely and the 
semimetallic state gives way to a fully gapped insulator.}
\label{fig:3}
\end{figure}

Thus, in the presence of the $\lambda \tau^y \sigma^z$ term, the two Weyl nodes are shifted in the opposite directions in energy, as seen in Fig.~\ref{fig:1}. 
It is no longer possible to align the Fermi energy with both nodes simultaneously and the semimetal at charge neutrality will have equal-volume 
electron and hole Fermi surfaces. The Weyl nodes exist as long as the electron and hole Fermi surfaces are separated. 
Upon increasing the magnitude of $\lambda$, the momentum-space separation between the nodes and between the electron and hole Fermi surfaces
decreases. Eventually, at a critical value $\lambda_{c1} = \sqrt{m^2 - m_{c1}^2}$, the Fermi surfaces touch and the Weyl nodes disappear, see Fig.~\ref{fig:2}. The Fermi surfaces still exist, however, 
until, at a still larger value of $\lambda_{c2} = \sqrt{(m+\Delta_D)^2 - \Delta_S^2}$, a fully gapped state is produced, see Fig.~\ref{fig:3}. 
In fact, the phase diagram of the system in the $m-\lambda$ plane is somewhat more involved, and we will explore it in more detail below. 

An important question, which arises in this context, is whether the topological properties of the Weyl semimetal, like the chiral edge states and the anomalous Hall conductivity, proportional
to the momentum-space separation between the Weyl nodes, survive when the nodes move away from the zero energy and the electron and hole Fermi surfaces appear. 
We will demonstrate below that this is indeed the case: the system retains its topological properties as long as it is gapless, in fact even after the Weyl nodes disappear above $\lambda = \lambda_{c1}$. 

To see this we start from the momentum-space Hamiltonian of the multilayer, which, after a canonical transformation:
\beq
\label{eq:21}
\sigma^{\pm} \rightarrow \tau^z \sigma^{\pm},\,\,\,\, \tau^{\pm} \rightarrow \sigma^z \tau^{\pm}, 
\eeq
takes the following form:
\beq
\label{eq;22}
{\cal H}(\bk) = v_F (\hat z \times \bsigma) \cdot \bk + [ m + \hat \Delta(k_z) ] \sigma^z + \lambda \tau^y.
\eeq
We now add a small external magnetic field $B$ along the $z$-direction (i.e. the growth direction of the multilayer), which we will
set back to zero at the end. 
The Hamiltonian becomes:
\beq
\label{eq:23}
{\cal H}(k_z) = v_F (\hat z \times \bsigma) \cdot \left( - i \bnabla + \frac{e}{c} \bA \right) + [ m + \hat \Delta(k_z) ] \sigma^z + \lambda \tau^y, 
\eeq
where $\bA = x B \hat y$ in the Landau gauge. 
To find the Landau levels, we introduce LL ladder operators in terms of the components of the kinetic momentum $\bpi = -i \bnabla + \frac{e}{c} \bA$ in the 
standard way (assuming $B > 0$ for concreteness):
\beq
\label{eq:24}
a = \frac{\ell_B}{\sqrt{2}}\left( \pi_x - i \pi_y \right), \,\,\,\, a^{\dag} = \frac{\ell_B}{\sqrt{2}}\left( \pi_x + i \pi_y \right), 
\eeq
where $\ell_B = \sqrt{c/ eB}$ is the magnetic length.
In terms of the ladder operators, the Hamiltonian takes the form:
\beq
\label{eq:25}
{\cal H}(k_z) = \frac{i \omega_B}{\sqrt{2}} \left( \sigma^+ a - \sigma^- a^{\dag} \right) +  [ m + \hat \Delta(k_z) ] \sigma^z + \lambda \tau^y, 
\eeq
where $\omega_B = v_F / \ell_B$. 
It is clear from Eq.~\eqref{eq:25}, that its eigenstates have the following general form:
\beqa
\label{eq:26}
|n \rangle&=&z_{n + \uparrow} |n-1,+, \uparrow \rangle + z_{n + \downarrow} |n,+, \downarrow \rangle + z_{n - \uparrow} |n-1,-, \uparrow \rangle \nonumber \\
&+&z_{n - \downarrow} |n,- , \downarrow \rangle,
\eeqa
where $|n,\pm, \uparrow\downarrow \rangle$ is the $n$th LL eigenstate on the top (+) or bottom (-) surface with spin up or down, and $z$ are complex amplitudes.  We have omitted all the additional eigenstate
labels on the left hand side for brevity, leaving only the LL index $n$. 
In the presence of the $\lambda \tau^y$ term, the above Hamiltonian can not be diagonalized analytically in general. 
However, we can find the $n = 0$ LL dispersions analytically, and this is in fact all that is needed for our purposes. 
Indeed, from Eq.~\eqref{eq:26} it is immediately clear that the $n = 0$ pair of LLs are polarized downwards, i.e. we can replace $\sigma^z = -1$
and then the $n = 0$ LL dispersion is given by the eigenvalues of:
\beq
\label{eq:27}
{\cal H}_{n = 0}(k_z) = - m - \hat \Delta(k_z) + \lambda \tau^y. 
\eeq
Diagonalizing Eq.~\eqref{eq:27}, we then find the $n = 0$ LL dispersions:
\beq
\label{eq:28}
\epsilon_{0 \pm}(k_z) = -m \pm \sqrt{\lambda^2 + \Delta^2(k_z) + 2 \lambda \Delta_D \sin(k_z d)}. 
\eeq
The $\epsilon_{0-}(k_z)$ level is below the Fermi energy $\epsilon_F = 0$ for all values of $k_z$. 
The $\epsilon_{0+}(k_z)$ LL, on the other hand, crosses the Fermi energy twice at the following momenta:
\beq
\label{eq:29}
k_z^{\pm} = \frac{\pi}{d} \pm  \frac{1}{d} \textrm{arccos} \left[\frac{f_{\mp}}{2 \Delta_D (\Delta_S^2 + \lambda^2)} \right], 
\eeq
where
\beqa
\label{eq:30}
f_{\pm}&=&\Delta_S  (\Delta_S^2 + \Delta_D^2 + \lambda^2 - m^2) \nonumber \\
&\pm&\lambda \sqrt{[\Delta_S^2 + \lambda^2 - (m - \Delta_D)^2] [(m + \Delta_D)^2 - \Delta_S^2 - \lambda^2]}. \nonumber \\
\eeqa 
This reduces to the already known result $k_z^{\pm} = \pi/d \pm k_0$,~\cite{Burkov11-1} where $k_0$ is given by Eq.~\eqref{eq:19}
in the limit $\lambda = 0$. 
As $\lambda$ increases, $k_z^-$ becomes equal to $\pi/d$ at the lower-critical value $\lambda_{c1} = \sqrt{m^2 - (\Delta_S - \Delta_D)^2}$, at which the 
electron and hole Fermi surfaces touch. At the upper critical value $\lambda_{c2} = \sqrt{(m + \Delta_D)^2 - \Delta_S^2}$, the Fermi surfaces disappear entirely, 
giving rise to a fully gapped state, and the $n = 0$ LL no longer crosses the Fermi energy. 

\begin{figure}[t]
\includegraphics[width=9cm]{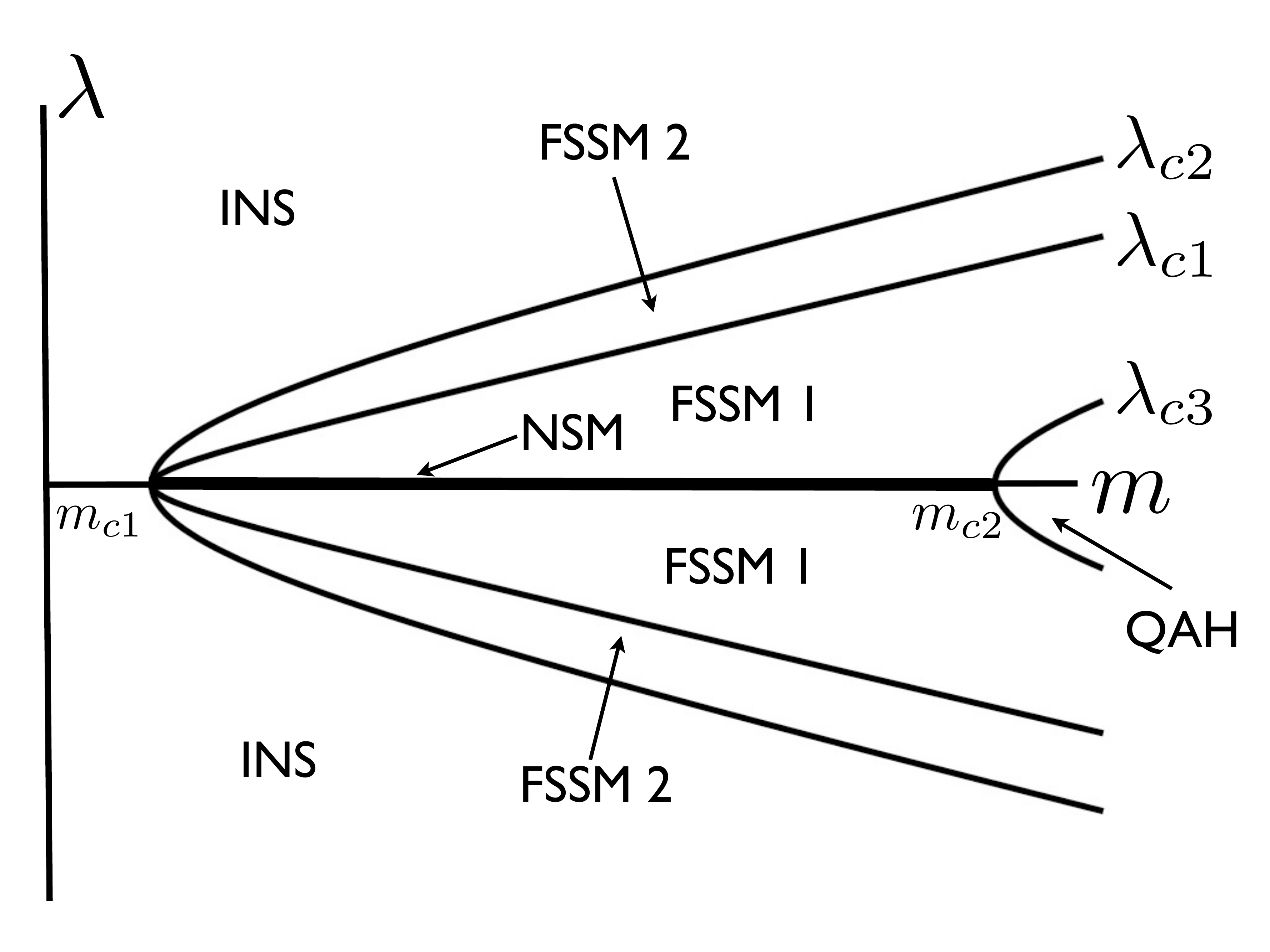}
\caption{Phase diagram of the multilayer structure in the $m-\lambda$ plane for $\Delta_S > \Delta_D$. There are five distinct phases: (1) Insulator (INS), outside of the $\lambda_{c2}$ line.  (2)  Nodal semimetal (NSM),
which exists along the interval $\lambda = 0$, $m_{c1} < m < m_{c2}$. (3) Semimetal with electron and hole pockets, but with Weyl nodes preserved (FSSM~1). This is bounded by the $\lambda_{c1} = 
\pm \sqrt{m^2 - m_{c1}^2}$ and $\lambda_{c3} = \pm \sqrt{(m-\Delta_D)^2 - \Delta_S^2}$ lines. (4) Semimetal without Weyl nodes (FSSM~2). This is bounded by the $\lambda_{c2} =  \pm \sqrt{(m+ \Delta_D)^2 - \Delta_S^2}$ and the 
$\lambda_{c1}$ lines. (5) QAH insulator phase, enclosed by the $\lambda_{c3}$ line. 
The anomalous Hall conductivity is nonzero in all semimetallic phases and the QAH phase.}
\label{fig:4}
\end{figure}

\begin{figure}[t]
\includegraphics[width=9cm]{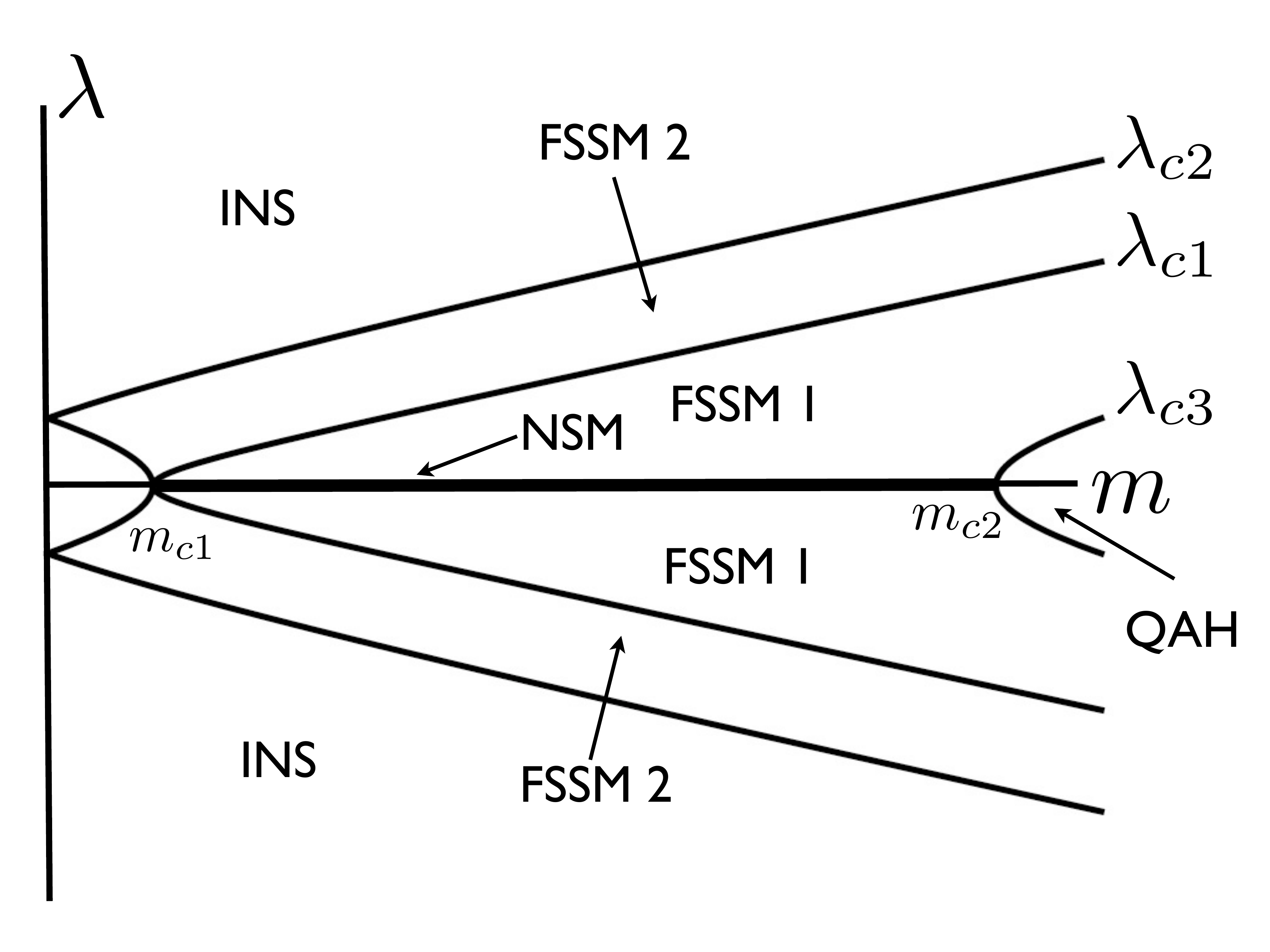}
\caption{Same as in Fig.~\ref{fig:4}, but for $\Delta_D > \Delta_S$. There is a additional transition in this case, which occurs as $\lambda$ is increased along the $m = 0$ line. 
The $|\lambda| < |\lambda_{c2}|$ phase is a TI, while the $|\lambda| > |\lambda_{c2}|$ phase is an ordinary insulator.}
\label{fig:5}
\end{figure}

There is a subtlety, associated with the $\lambda_{c2}(m)$ critical line. Two cases need to be considered separately. When $\Delta_{S} > \Delta_D$, i.e. when the multilayer in the presence 
of both TR and I is an ordinary insulator, the $\lambda_{c2}(m)$ line starts at $m = m_{c1}$ and separates a semimetal from an ordinary insulator, as shown in Fig.~\ref{fig:4}. 
When $\Delta_D > \Delta_S$, however, the system is a topological insulator when TR and I are unbroken. If TR is preserved, but I is broken by increasing the magnitude of $\lambda$, there
is eventually a TI to NI transition at a critical value of $\lambda = \sqrt{\Delta_D^2 - \Delta_S^2}$. Correspondingly, the $\lambda_{c2}(m)$ line has a different shape, shown in Fig.~\ref{fig:5}. 

Finally, a third critical line $\lambda_{c3} = \sqrt{(m - \Delta_D)^2 -\Delta_S^2}$ exists, separating 3D quantum anomalous Hall (QAH) insulator~\cite{Burkov11-1} from the semimetallic phase with Weyl nodes 
and Fermi surfaces.  

The most important consequence of the above analysis is that the anomalous Hall conductivity:
\beq
\label{eq:32}
\sigma_{xy} = \int_{k_z^-}^{k_z^+} \frac{d k_z}{2 \pi} \frac{e^2}{h} = \frac{e^2 (k_z^+ - k_z^-)}{2 \pi h}, 
\eeq
is nonzero beyond $\lambda = \lambda_{c1}$, when the Weyl nodes disappear, persisting all the way 
to the $\lambda = \lambda_{c2}$ point, when a fully gapped insulating state emerges (chiral surface states disappear when the Weyl nodes disappear at $\lambda = \lambda_1$).  
This follows simply from the fact that one of the two $n = 0$ LLs always dips below the Fermi energy
in the interval $k_z^- < k_z < k_z^+$ (the second $n = 0$ LL is always below the Fermi energy)
in all gapless phases, thus giving rise to a nonzero Hall conductivity (relative to the insulating state at $m < m_{c1}$, where there is one zero-mode LL above and one zero-mode LL below the Fermi 
energy and the Hall conductivity is zero).~\cite{Burkov11-1} The connection between the nonzero anomalous Hall conductivity and the zero-mode LL crossing of the 
Fermi energy can be seen most easily by referring to the St\v{r}eda formula for the quantized nonclassical 
part of the anomalous Hall conductivity:~\cite{Streda82,Haldane88}
\beq
\label{eq:32a}
\sigma_{xy} = - c \left.\frac{\partial \delta N}{\partial B} \right|_{B=0}, 
\eeq
where $\delta N$ is the extra charge density, corresponding to the filled zero-mode LL, when $k_z^- < k_z < k_z^+$: 
\beq
\label{eq:32b}
\delta N = - \frac{e}{2 \pi \ell_B^2}  \int_{k_z^-}^{k_z^+} \frac{d k_z}{2 \pi} = - \frac{e^2 (k_z^+ - k_z^-) B}{2 \pi h c}. 
\eeq
The nonquantized part of the anomalous Hall conductivity, on the other hand, is always zero due to charge neutrality as long as the Fermi level remains pinned at $\epsilon_F  = 0$.  
The full phase diagram of the multilayer in the $m - \lambda$ plane is shown in Figs.~\ref{fig:4}, \ref{fig:5}. 

\section{Magnetic field driven equilibrium current}
\label{sec:4}
In this section we will demonstrate that, in addition to the semi-quantized AH conductivity and chiral ``Fermi-arc" surface state, the TI-NI multilayer 
with broken TR and inversion is characterized by another topological phenomenon, namely the external magnetic field driven persistent current, 
which we have introduced in section~\ref{sec:2} using general arguments, based on Chern-Simons like description of the Weyl semimetal.
Here we will limit our analysis to small values of $\lambda$, in which case all calculations can be done analytically, and which is perhaps
the most relevant regime experimentally.  

We will calculate the equilibrium current in linear response with respect to the $\lambda \tau^y \sigma^z$ term. In this case, the 
energy separation between the Weyl modes:
\beq
\label{eq:33}
\Delta \epsilon = | \Delta \epsilon_- - \Delta \epsilon_+ | = \frac{\lambda}{\Delta_S |m|} \sqrt{(m_{c2}^2 - m^2) (m^2 - m_{c1}^2)}, 
\eeq
can be thought of as an ``electrochemical potential difference" between them. We thus want to calculate an electrical current 
that arises in response to this ``electrochemical potential difference". Note that we have defined $\Delta \epsilon$ as the energy 
shift of the node of positive chirality (which is at $k_z^-$ for $m > 0$ but at $k_z^+$ for $m < 0$) minus the energy shift of the node of negative chirality: defined this way, $\Delta \epsilon$ is 
independent of the sign of $m$. 
We also assume that an external magnetic field of magnitude $B$ is applied along the growth direction of the multilayer, which we take to be 
the $z$-direction (note that the spin magnetization $m$ does not contribute to the total internal orbital magnetic field for a sample of an infinite 
slab geometry, due to perfect cancellation with the corresponding demagnetizing field). We will ignore the Zeeman contribution of the external field, 
assuming it to give only a small correction to $m$. 
The unperturbed Hamiltonian, after the canonical transformation of Eq.~\eqref{eq:21}, is given then by:
\beq
\label{eq:34}
{\cal H}(k_z) = v_F (\hat z \times \bsigma) \cdot \left( - i \bnabla + \frac{e}{c} \bA \right) + [ m + \hat \Delta(k_z) ] \sigma^z. 
\eeq
Diagonalizing $\hat \Delta(k_z)$, which is a constant of motion, brings the Hamiltonian to a block-diagonal form, with two independent 
$2 \times 2$ blocks:
\beq
\label{eq:35}
{\cal H}_{\pm}(k_z) = v_F (\hat z \times \bsigma) \cdot \left( - i \bnabla + \frac{e}{c} \bA \right) + m_{\pm}(k_z) \sigma^z,
\eeq
where $m_{\pm}(k_z) = m \pm \Delta(k_z)$. 
Using Landau gauge $\bA = x B \hat y$ (for convenience, we will implicitly assume $B > 0$ in all calculations, but the final result does depend on the sign of $B$) 
and introducing LL ladder operators as in section~\ref{sec:3}, we obtain:
\beq
\label{eq:36}
{\cal H}_{\alpha}(k_z) = \frac{i \omega_B}{\sqrt{2}} ( \sigma^+ a - \sigma^- a^{\dag} ) + m_{\alpha}(k_z) \sigma^z,
\eeq
where $\alpha = \pm$. 
It is clear that the eigenstates of each $2 \times 2$ block of Eq.~\eqref{eq:36} have the following general form:
\beq
\label{eq:37}
| v^{n s \alpha} \rangle  = v^{n s \alpha} _{\uparrow} | n-1, \uparrow \rangle + v^{n s \alpha}_{\downarrow} |n, \downarrow \rangle. 
\eeq
Here $n = 0,1,2,\ldots$ is the LL index and $s = \pm$ labels the electron-like and hole-like LLs correspondingly.
The eigenvalues of Eq.~\eqref{eq:36} are easily found and are given by:
\beqa
\label{eq:38}
\epsilon_{n s \alpha} (k_z)&=&s \sqrt{2 \omega_B^2 n + m_{\alpha}^2(k_z)}, \,\, n \ge 1, \nonumber \\
\epsilon_{0 \alpha}&=&- m_{\alpha}(k_z). 
\eeqa
The corresponding eigenvectors are:
\beqa
\label{eq:39}
| v^{s \alpha}_n \rangle&=&\frac{1}{\sqrt{2}}\left(\sqrt{1 + \frac{m_{\alpha}(k_z)}{\epsilon_{n s \alpha}}}, i s \sqrt{1 - \frac{m_{\alpha}(k_z)}{\epsilon_{n s \alpha}}} \right), \,\, n \ge1, \nonumber \\
v^{\alpha}_0&=&(0, 1). 
\eeqa
We note the following symmetry properties of the eigenfunctions Eq.~\eqref{eq:39}, which will play an important role in our analysis:
\beq
\label{eq:39a}
\langle v^{+ \alpha}_n | v^{- \alpha'}_n \rangle = - \langle v^{+ \alpha'}_n | v^{- \alpha}_n \rangle, \,\,\, n \ge 1,
\eeq
while:
\beq
\label{eq:39b}
\langle v^{+ \alpha}_n | \sigma^z |  v^{- \alpha'}_n \rangle = \langle v^{+ \alpha'}_n | \sigma^z | v^{- \alpha}_n \rangle, \,\,\, n \ge 1.
\eeq

The total four-component spinor eigenfunctions of Eq.~\eqref{eq:36}, which we will denote simply by  $| n s \alpha \rangle$, are given by the tensor product of $| v^{s \alpha}_n \rangle$ and the eigenfunctions of the $\hat \Delta(k_z)$ operator:
\beq
\label{eq:40}
|u^{\alpha} \rangle = \frac{1}{\sqrt{2}} \left( 1, \alpha \frac{\Delta_S + \Delta_D e^{- i k_z d}}{\Delta(k_z)} \right), 
\eeq
i.e.
\beq
\label{eq:41}
|n s \alpha \rangle = | u^{\alpha} \rangle \otimes | v^{s \alpha}_n \rangle. 
\eeq   
Let us point out the following matrix elements of the eigenstates of the tunneling operator, which will be important in the calculation below:
\beqa
\label{eq:40a}
\langle u^+ | \tau^x | u^- \rangle&=&i \frac{\Delta_D \sin(k_z d)}{\Delta(k_z)}, \nonumber \\
\langle u^+ | \tau^y | u^- \rangle&=&i \frac{\Delta_S + \Delta_D \cos(k_z d)}{\Delta(k_z)}. 
\eeqa

We evaluate the electrical current in the $z$-direction in linear response with respect to the inversion-symmetry violating term $\lambda \tau^y \sigma^z$. 
The standard expression is: 
\beqa
\label{eq:42}
j_z &=& \frac{\lambda}{2 \pi \ell_B^2 L_z} \sum_{n, n', s, s', \alpha, \alpha', k_z} \langle n s \alpha | \hat j_z | n' s' \alpha'  \rangle  \nonumber \\
&\times&\langle n' s' \alpha' | \tau^y | n s \alpha \rangle
\frac{n_F(\epsilon_{n s \alpha}) - n_F(\epsilon_{n' s' \alpha'})}{\epsilon_{n s \alpha} - \epsilon_{n' s' \alpha'}}, 
\eeqa
 where $L_z$ is the size of the system in the growth direction, we have suppressed explicit dependence on $k_z$ everywhere for brevity, and we are 
 using the momentum space Hamiltonian Eq.~\eqref{eq:34}, i.e. after the canonical transformation of Eq.~\eqref{eq:21}.   
The current operator $\hat j_z$ is given by:
\beq
\label{eq:43}
\hat j_z = - e \frac{\partial {\cal H}(k_z)}{\partial k_z} = e \Delta_D d \sigma^z \left[ \tau^x \sin(k_z d) + \tau^y \cos(k_z d) \right]. 
\eeq
To evaluate Eq.~\eqref{eq:42}, we first note that both $\hat j_z$ and $\tau^y$ operators are diagonal in the spin indices, which means, taking into account 
Eq.~\eqref{eq:37}, that the matrix elements of these operators, appearing in Eq.~\eqref{eq:42}, are diagonal in the LL index $n$. 
Furthermore, if the Fermi level is pinned by charge neutrality at zero energy $\epsilon_F = 0$, only the terms with $s \ne s'$ contribute to Eq.~\eqref{eq:42}
due to the difference of the Fermi distribution functions in the numerator (we assume the temperature to be lower than all other energy scales in the problem). 
Then we obtain:
\beqa
\label{eq:44}
j_z &=&\frac{\lambda}{2 \pi \ell_B^2 L_z} \sum_{n, s \ne s', \alpha, \alpha', k_z} \langle n s \alpha | \hat j_z | n s' \alpha'  \rangle  \nonumber \\
&\times&\langle n s' \alpha' | \tau^y | n s \alpha \rangle
\frac{n_F(\epsilon_{n s \alpha}) - n_F(\epsilon_{n s' \alpha'})}{\epsilon_{n s \alpha} - \epsilon_{n s' \alpha'}}, 
\eeqa
To simplify this further we use Eqs.~\eqref{eq:39a}, \eqref{eq:39b} and \eqref{eq:40a}. According to these, the product of the matrix elements of $\hat j_z$ and $\tau^y$, appearing in Eq.~\eqref{eq:44}, 
is antisymmetric with respect to the interchange of the $\alpha$ and $\alpha'$ indices for all $n \ge 1$. This means that the contribution of all $n \ge 1$ LLs cancels out and only the two 
$n = 0$ LLs contribute to the expectation value of the current. 
Then we obtain:
\beqa
\label{eq:45}
j_z &=&\frac{\lambda}{2 \pi \ell_B^2 L_z} \sum_{\alpha \ne \alpha', k_z} \langle 0 \alpha | \hat j_z | 0 \alpha'  \rangle \langle 0 \alpha' | \tau^y | 0 \alpha \rangle \nonumber \\
&\times&\frac{n_F(\epsilon_{0 \alpha}) - n_F(\epsilon_{0 \alpha'})}{\epsilon_{0 \alpha} - \epsilon_{0 \alpha'}}.
\eeqa
Since $\epsilon_{0+}(k_z) = - m_+(k_z) = -m - \Delta(k_z)$ is always negative, while $\epsilon_{0-}(k_z) = -m_-(k_z) = -m + \Delta(k_z)$ changes sign at the locations 
of the Weyl nodes, only the subset of the 1D BZ, in which $m_+(k_z)$ and $m_-(k_z)$ have opposite signs, contributes to Eq.~\eqref{eq:45}. 
Evaluating the matrix elements in Eq.~\eqref{eq:45}, we then obtain:
\beqa
\label{eq:46}
j_z&=&\frac{ e \Delta_D d}{2 \pi \ell_B^2} \int_{\pi/d - k_0}^{\pi/d + k_0} \frac{d k_z}{2 \pi} \nonumber \\
&\times&\frac{\Delta^2(k_z) \cos(k_z d) + \Delta_S \Delta_D \sin^2(k_z d)}{\Delta^3(k_z)}, 
\eeqa
where we have used the fact that the above integral, evaluated over the whole BZ, vanishes. 
Evaluating the integral over the interval between the locations of the Weyl nodes in Eq.~\eqref{eq:47}, we obtain:
\beq
\label{eq:47}
j_z = - \frac{e^2 B \lambda}{4 \pi^2 | m | \Delta_S c} \sqrt{(m_{c2}^2- m^2) (m^2 - m_{c1}^2)}. 
\eeq
Using Eq.~\eqref{eq:33} for the energy difference between the Weyl nodes, induced by the inversion-breaking term $\lambda \tau^y \sigma^z$, we 
finally obtain:
\beq
\label{eq:48}
j_z = - \frac{e^2 \Delta \epsilon}{4 \pi^2 \hbar^2 c} B, 
\eeq
where we have restored explicit $\hbar$. Note that $B$ in Eq.~\eqref{eq:48} can have both signs, and the current correctly changes sign under TR, since 
$\Delta \epsilon$ is TR-invariant. 
There is thus indeed a purely equilibrium nondissipative current, as predicted by the Chern-Simons theory analysis of section~\ref{sec:2}. 
The current is proportional to both the applied external field and to energy separation (for small separation) between the Weyl nodes. 
The remaining coefficient is a universal combination of fundamental constants.

The above result can be understood physically by appealing to the well-known Adler-Bell-Jackiw anomaly.~\cite{Adler69,Jackiw69,Nielsen83} 
Following the argument by Nielsen and Ninomiya,~\cite{Nielsen83,Kharzeev08} let us imagine that there is an electric field $E$, applied in the same direction as the magnetic 
field $B$. In the presence of the magnetic field, the electric field will lead to transfer of particles between the two Weyl nodes through the zero-mode LL, which 
has a definite chirality near each of the Weyl nodes, reflecting the chirality of the nodes. The rate of the particle transfer between the left ($-$) and right ($+$) node per unit volume is given by:~\cite{Nielsen83}
\beq
\label{eq:49}
\frac{d (N_- - N_+)}{d t} =  2\,\, \frac{e E}{2 \pi \hbar} \,\, \frac{1}{2 \pi \ell_B^2} =  \frac{e^2}{2 \pi^2 \hbar^2 c } E B. 
\eeq 
Since the Weyl nodes have an energy difference $\Delta \epsilon$, this particle transfer process has an associated power cost per unit volume:
\beq
\label{eq:50}
P = \Delta \epsilon \,\,\frac{d (N_- - N_+)/2}{d t}, 
\eeq
This power is provided by the current of magnitude $j$, so that $P = j E$. Thus we obtain:
\beq
\label{eq:51}
j E =  \Delta \epsilon \frac{e^2}{4 \pi^2 \hbar^2 c } E B. 
\eeq
Cancelling $E$ on both sides of this equation and sending it to zero, we obtain:
\beq
\label{eq:52}
j =  \frac{e^2 \Delta \epsilon}{4 \pi^2 \hbar^2 c} B, 
\eeq
which is identical to Eq.~\eqref{eq:48} (the sign of the current is not determined by the above energy balance argument).  

\section{Discussion and conclusions}
Let us now address the experimental observability of the proposed magnetic-field-driven persistent current. 
First let us estimate possible current density magnitudes. We take the energy separation between the Weyl nodes 
$\Delta \epsilon \sim 1$~meV and the magnetic field $B \sim 1$~T. This gives $j_z \sim 0.1~\textrm{A/cm}^2$. This is 
certainly an easily measurable current.  

The effect itself will of course exist only at low enough temperatures and in clean enough samples, so that, in particular, 
the temperature and the impurity scattering rate are smaller than $\Delta \epsilon$, the energy separation between the Weyl nodes. 
This implies temperatures less than about $10$~K, if we take $\Delta \epsilon \sim 1$~meV. 
Thus the proposed effect may be observable. The biggest unknown here is of course the value of the spin-orbit interaction 
parameter $\lambda$. In the above estimate we have assumed it to be of order $\sim 1$~meV. It may, however, turn out to be 
much smaller, in which case the effect we have described may turn out to be unobservable. 

In conclusion, we have studied Weyl semimetal, realized in a TI-NI multilayer heterostructure, in which both TR and I symmetries are violated. 
Breaking TR symmetry moves the Weyl nodes to different points in momentum space, while breaking inversion shifts them to different energies, 
thus producing a state with equal-volume electron and hole pockets at charge neutrality. We have demonstrated that if an external magnetic field 
is applied to this system along the direction of the vector, connecting the Weyl nodes, an equilibrium nondissipative current flows in this system. 
The current is proportional to both the applied magnetic field and the energy difference between the Weyl nodes, with a coefficient that is a universal 
combination of fundamental constants. This current may be measurable under experimentally-achievable conditions.  
In a sample, not connected to external leads, a voltage in response to an applied magnetic field will instead be measured. This effect can potentially be useful 
for magnetic field detection. 
\label{sec:5} 
\begin{acknowledgments}
We acknowledge useful discussions with Sung-Sik Lee. Financial support was provided by the NSERC of Canada and a University of Waterloo start-up grant. 
\end{acknowledgments}


\begin{thebibliography}{99}
\bibitem{footnote1} Strictly speaking, this type of topological protection is imperfect, in the sense that node mixing is possible 
by a perturbation, which has a nonzero Fourier component at the momentum, separating the Weyl nodes. In practice, this is not a 
significant restriction. The main type of perturbation, that will be present in experimental systems, in disorder potential. This will, in general, 
have a nonzero Fourier component at any momentum. However, as long as this Fourier component has a zero mean, as one would expect 
for a random potential, it will not open a gap.
\end{thebibliography}
\end{document}